\input{aipcheck}

\documentclass[
    ,final            
  ]
  {aipproc}

\layoutstyle{8x11single}

\begin{document}

\title{POET: POlarimeters for Energetic Transients}


\author{J. E. Hill}{
  address={Universities Space Research Association, CRESST and NASA Goddard Space Flight Center, Greenbelt, MD 20771 USA}
}
\author{M. L. McConnell}{
address={Space Science Center, University of New Hampshire, Durham, NH 03824, USA}
}
 \author{P. Bloser}{
 address={Space Science Center, University of New Hampshire, Durham, NH 03824, USA}
 }
\author{J. Legere}{
address={Space Science Center, University of New Hampshire, Durham, NH 03824, USA}
}
\author{J. Macri}{
address={Space Science Center, University of New Hampshire, Durham, NH 03824, USA}
}
\author{J. Ryan}{ 
 	address={Space Science Center, University of New Hampshire, Durham, NH 03824, USA}
}
\author{S. Barthelmy}{
address={NASA Goddard Space Flight Center, Greenbelt, MD 20771, USA}
}
\author{L. Angelini}{
address={NASA Goddard Space Flight Center, Greenbelt, MD 20771, USA}
}
\author{T. Sakamoto }{
	address={NASA Goddard Space Flight Center, Greenbelt, MD 20771, USA}
}
\author{J. K. Black}{
	address={Rock Creek Scientific, 1400 East-West Hwy, Suite 807, Silver Spring, MD 20910, USA}
}
\author{D. H. Hartmann}{ 
	address={Department of Physics and Astronomy, Clemson University, Clemson, SC 29634}
}
\author{P. Kaaret}{ 
	address={Department of Physics and Astronomy, University of Iowa, Iowa City, IA 52242, USA}
}
\author{B. Zhang}{ 
	address={Department of Physics, University of Nevada, Box 454002, Las Vegas, NV 89154, USA}
}
\author{K. Ioka}{
address={Theoretical Astrophysics Group, Department of Physics, Kyoto University, Sakyo-ku, Kyoto 606-8502, Japan}
}
\author{T. Nakamura}{
address={Theoretical Astrophysics Group, Department of Physics, Kyoto University, Sakyo-ku, Kyoto 606-8502, Japan}
}
\author{K. Toma}{ 
	address={Theoretical Astrophysics Group, Department of Physics, Kyoto University, Sakyo-ku, Kyoto 606-8502, Japan}
}
\author{R. Yamazaki}{ 
	address={Department of Physical Science, Hiroshima University, 
Higashi-Hiroshima, 739-8526, Japan}
}
 \author{X. Wu}{ 
	address={Purple Mountain Observatory, Chinese Academy of Sciences, Nanjing 210008,
China}
}

\begin{abstract}
POET (Polarimeters for Energetic
Transients) is a Small Explorer mission concept proposed to NASA in January 2008. The principal
scientific goal of POET is to measure GRB
polarization between 2 and 500 keV. The payload
consists of two wide FoV instruments: a Low
Energy Polarimeter (LEP) capable of polarization
measurements in the energy range from 2-15
keV and a high energy polarimeter (Gamma-Ray
Polarimeter Experiment -- GRAPE) that will
measure polarization in the 60-500 keV energy
range. Spectra will be measured from 2 keV up
to 1 MeV. The POET spacecraft provides a
zenith-pointed platform for maximizing the
exposure to deep space. Spacecraft rotation will
provide a means of effectively dealing with
systematics in the polarization response. POET
will provide sufficient sensitivity and sky
coverage to measure statistically significant
polarization for up to 100 GRBs in a two-year mission.
Polarization data will also be obtained for solar
flares, pulsars and other sources of astronomical
interest.
\end{abstract}

\maketitle


\section{Introduction}

Gamma-ray bursts (GRBs) are the most explosive events in the universe,
and have stimulated intense observational and theoretical research.
Theoretical models indicate that an understanding of the
inner structure of GRBs, including the geometry and physical
processes close to the central engine, requires the exploitation
of high energy X-ray and gamma-ray polarimetry but observational techniques have been limited. Recent advances in instrument capabilities finally enable the
exploration of polarization of X-ray and gamma-ray emissions
from GRBs. POlarimeters for Energetic Transients (POET) is a SMEX mission concept that provides  highly sensitive polarimetric
observations of GRBs and can also make polarimetry
measurements of solar flares, pulsars, soft gamma repeaters,
and slow transients.
POET will make measurements
with two different polarimeters (both with wide fields of view) to provide a
broad energy range of observations:
The Gamma-RAy Polarimeter Experiment (GRAPE; 60--500 kev) and the
Low Energy Polarimeter (LEP; 2--15 keV).
The POET mission would significantly advance
our understanding of key physical processes through high
energy polarimetry.

\section{Gamma-Ray Burst Science With POET}
Extensive multi-wavelength observations of the prompt GRB emission
and the long-wavelength afterglows have led to the development
of models for two types of GRBs \citep{ZM2004}. Long bursts (> 2
s) typically exhibit a relatively soft spectrum and are generally
associated with the death of massive stars\citep{Pac1998} whereas short bursts (< 2 s) generally
exhibit a harder spectrum and are associated with the
merger of compact star binaries (e.g., neutron star-neutron
star, neutron star-black hole, etc., \citep{Gehrels05}.
In both cases, the process is thought to result in the formation
of a black hole. Regardless of the progenitor and the central
engine, a generic "fireball" model suggests that a
relativistic jet is launched from the center of the explosion, with
a bulk Lorentz factor, $\Gamma$, greater than 100 \citep{RM1992}. 
The "internal" dissipation within the fireball (likely due
to internal shocks) leads to emission in the X-ray and gamma-ray
band, which corresponds to the observed GRB prompt
emission. Eventually, the fireball jet is decelerated by the
circumburst medium, which leads to a long-lasting forward
shock, the emission from which is believed to be responsible for
the long-lasting afterglows following the bursts themselves.

In spite of extensive observational efforts (e.g., CGRO, HETE-2,
BeppoSAX, INTEGRAL, {\it  Swift}, and others), several key
properties of GRB explosions remain poorly understood and are
difficult or even impossible to infer with the spectral and lightcurve
information currently collected. High energy polarization measurements
will lead to unambiguous answers to many open questions, including:

\begin{description}
\item[What is the magnetic structure of GRB jets?] 
It is speculated that strong magnetic fields are generated at the GRB central
engine, which may play an essential role in the launch of the
relativistic jets. It is unclear, however, whether the GRB emission
region is penetrated by a globally structured, dynamically
important magnetic field, and whether GRB emission is due to
shock dissipation or magnetic reconnection.

\item[What is the geometric structure of GRB jets?]
Although it is
generally believed that GRBs are collimated, the distribution
of jet opening angles and the observer's viewing direction are
not known, and it is not clear whether there are small-scale
structures within the global jet.

\item[What is the prompt radiation mechanism of GRBs?] The
leading model is synchrotron emission of relativistic electrons
in a globally ordered magnetic field carried from the central
engine or random magnetic fields generated in situ in the shock
dissipation region. Other suggestions include Compton drag of
ambient soft photons, synchrotron self-Compton emission, and
the combination of a thermal component from the photosphere
and a non-thermal component (e.g., synchrotron).
\end{description}

The theoretical models for the prompt emission generally fall into two broad types \citep{Wax03}: \textbf{The physical model} invokes a globally ordered magnetic field
in the emission region, so that electron synchrotron emission in
this ordered field gives a net linear polarization (e.g., \citep{Wax03}\citep{LPB2003}\citep{Gran03}). Such a model applies
for most observer viewing-angle geometries, where the typical
level of polarization is quite high ($\Pi >20 \%$), with the maximum 
$\Pi$ $\sim 70\%$. \textbf{The geometric model} requires an optimistic viewing direction
to observe a high degree of polarization. In this model, both
the magnetic field structure and electron energy distribution
is random in the emission region so that no net polarization is
detected if the viewing angle is along the jet beam (regardless
of the radiation mechanism). However, if the viewing direction is
near the edge of the jet, in particular about 1/$\Gamma$ outside the jet
cone, a high polarization degree would result due to loss of the
emission symmetry \citep{SD1995}\citep{Wax03}\citep{Laz04}. Within 
the context of the geometric model, typical
polarization values are $\Pi <20\%$, although synchrotron emission
can produce as high as $\Pi \sim70\%$ and Compton drag models
\citep{Laz04} can achieve $\Pi \sim100\%$ under optimistic
geometric configurations.

In general, given a random distribution of viewing angles, the
fraction of bursts that can achieve a high $\Pi$ in the geometric
models is significantly smaller than that in the physical models.
A statistical study of polarization properties of a large sample
of GRBs would therefore differentiate between the models, and
provide a direct diagnostic of the magnetic field structure,
radiation mechanism and geometric configuration of GRB jets.
Toma et al. \cite{Toma08} displays the predictions for
the distribution of polarization magnitudes and the dependence
of $\Pi$ on the peak of the energy break ($E_p$) in the GRB spectrum.
Results are shown for three distinct models:
\begin{itemize}
\item Physical model for synchrotron emission with ordered
magnetic fields (SO)
\item Geometric model for synchrotron emission in random
magnetic fields (SR)
\item Geometric model for Compton-drag (CD)
\end{itemize}

The POET sensitivity will allow the accumulation of GRB polarization measurements at a
rate of $\sim$ 50/year, permitting studies that will distinguish between
the geometric and physical models. Given a sufficiently large
number of events, it may even be possible to distinguish
between the two geometric models (SR and CD).

The GRB radiation mechanism can be determined from the
energy-dependence of the polarization measurements. The GRB prompt
emission spectrum is typically characterized by a broken powerlaw
\citep{Band93}. Generally it is
believed that spectral energy break ($E_p$) corresponds to a break
in the non-thermal electron spectral distribution. This would
result in a jump of the polarization degree across $E_p$. Alternatively, some authors (e.g., \citep{Ryde05}) argue that the observed spectrum is a superposition of a
thermal spectrum (probably from the photosphere) and a non-thermal 
synchrotron spectrum. This would lead to a dip in the
polarization spectrum. POET can measure the
degree of polarization both above and below $E_p$ (or measure
the polarization as a function of energy) and therefore would differentiate the
two models, and identify the GRB radiation mechanism
and the emission site, which are very difficult to infer from
current observations.

\section{POET Instrument Suite}


\begin{table}
\begin{tabular}{lll}
\hline
&\tablehead{1}{l}{b}{GRAPE}
& \tablehead{1}{l}{b} {LEP}
 \\
\hline
Polarimetry    & 60--500 keV & 2--15 keV\\
Detectors & BGO/plastic scintillator (62) & $Ne:CO_2:CH_3NO_2$ Gas (8)\\
Spectroscopy &  15 keV -- 1 MeV & 2 -- 15 keV\\
Detectors & NaI(TI) scintillator (2) & as above\\
Field-of-View & $\pm60^o$ & $\pm44^o$  \\
\hline
\end{tabular}
\caption{Instrument Parameters}
\label{tabparam}
\end{table}

POET is comprised of two polarimetry instruments, GRAPE and LEP, 
co-aligned on a zenith pointed rotating spacecraft.
LEP and GRAPE determine polarization by measuring
the number of events versus the event azimuth angle (EAA) as
projected onto the sky. This is referred to as a modulation profile
and represents a measure of the polarization magnitude and
direction of polarization for the incident beam. Depending on the
type of polarimeter, the EAA is either the direction of the ejected
photoelectron (LEP) or the direction of the scattered photon
(GRAPE). The response of a polarimeter to $100\%$ polarized
photons can be quantified in terms of the modulation factor, $\mu$,
which is given by:
\begin{equation}
\mu=\frac{C_{max}-C_{min}}{C_{max}+C_{min}}  \label{mod}
\end{equation}
Where $C_{max}$ and $C_{min}$ are the maximum and minimum of the
modulation profile, respectively. The polarization fraction
($\Pi$) of the incident flux is obtained by dividing the measured
modulation by that expected for $100\%$ polarized flux. The
polarization angle ($\phi_o$) corresponds either to the maximum of
the modulation profile (LEP) or the minimum of the modulation
profile (GRAPE). To extract these parameters from the data, the
modulation histograms are fit to the functional form:
\begin{equation}
C(\Phi)=A+B\cos^2(\Phi-\Phi_o)  \label{C_phi}
\end{equation}

The sensitivity of a polarimeter is defined in terms of the
MDP, which refers to the minimum level of polarization that
is detectable with a given observation (or, equivalently, the
apparent polarization arising from statistical fluctuations in
unpolarized data). The precise value of the MDP will depend
on the source parameters (fluence, spectrum, etc.) and the
polarimeter characteristics. At the $99\%$ confidence level, the
MDP can be expressed as,
\begin{equation}
MDP=\frac{4.29}{\epsilon\mu} {(\frac{\epsilon S A+B}{t})}^{1/2} \label{MDP}
\end{equation}
where S is the source strength (cts $cm^{-2} s^{-1}$), B is the total
background rate (cts $cm^{-2}s^{-1}$), t is the observing time (sec), $\epsilon$
is the quantum efficiency, and A is the collecting area. The
ultimate sensitivity, however, may not be limited by statistics
but by systematic errors created by false modulations that arise
from azimuthal asymmetries in the instrument.


\subsection{Gamma-RAy Polarimeter Experiment: GRAPE}

GRAPE \citep{McCon04}\citep{McCon08} is designed to measure polarization from 60-500 keV and to provide spectroscopy over a
broad energy range from 15 keV to 1 MeV. The GRAPE design
is highly modular and fault tolerant. The GRAPE instrument is
composed of 64 independent detector modules arranged in two
identical assemblies that provide the associated electronics
and the required mechanical and thermal support. All
detector modules employ well-established, high-technology
readiness level scintillator/photomultiplier tube (PMT)
technology with geometries optimized by both simulation and
laboratory studies.

At energies from $\sim50$ keV up to several MeV, photon interactions
are dominated by Compton scattering. The operational concept
for GRAPE is based on the fact that, in Compton scattering,
photons are preferrentially scattered at a right angle to the
incident electric field vector (the polarization vector). If the
incident beam of photons is polarized, the azimuthal distribution
of scattered photons will be asymmetric.  The
direction of the polarization vector is defined by the minimum
of the scatter angle distribution. The GRAPE performance
characteristics are shown in Table \ref{tabparam}.

\begin{figure}
  \includegraphics[height=.2\textheight]{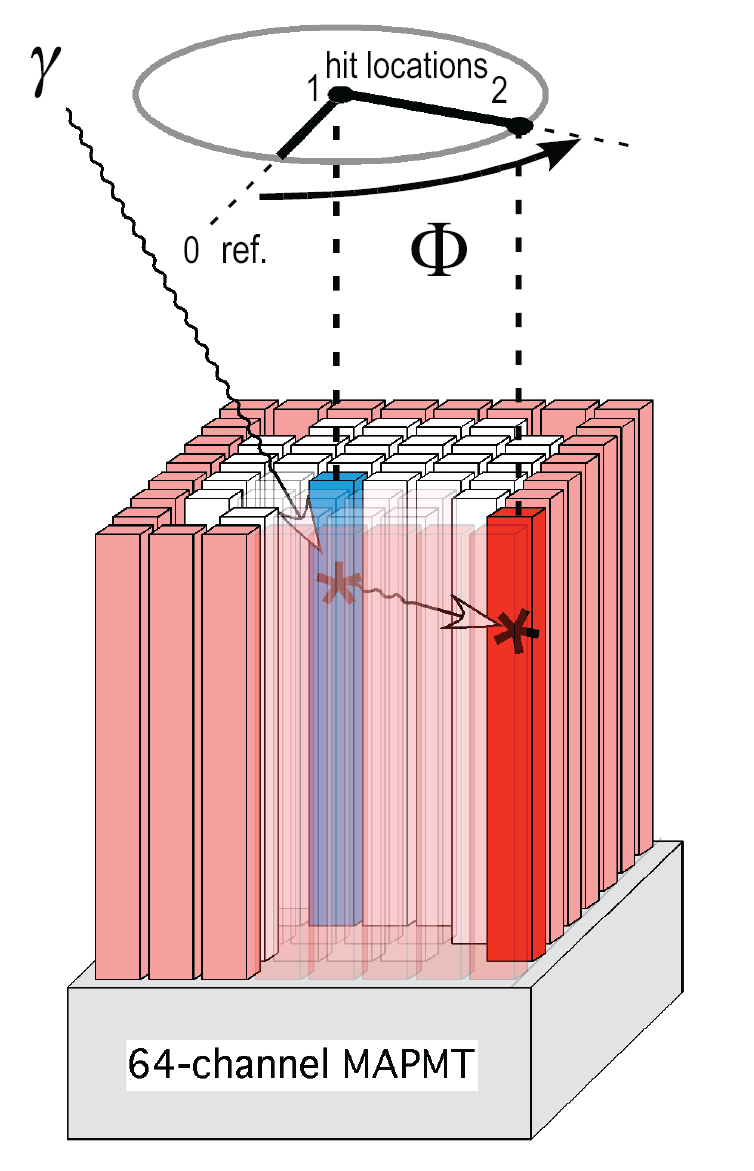}
    \includegraphics[height=0.2\textheight]{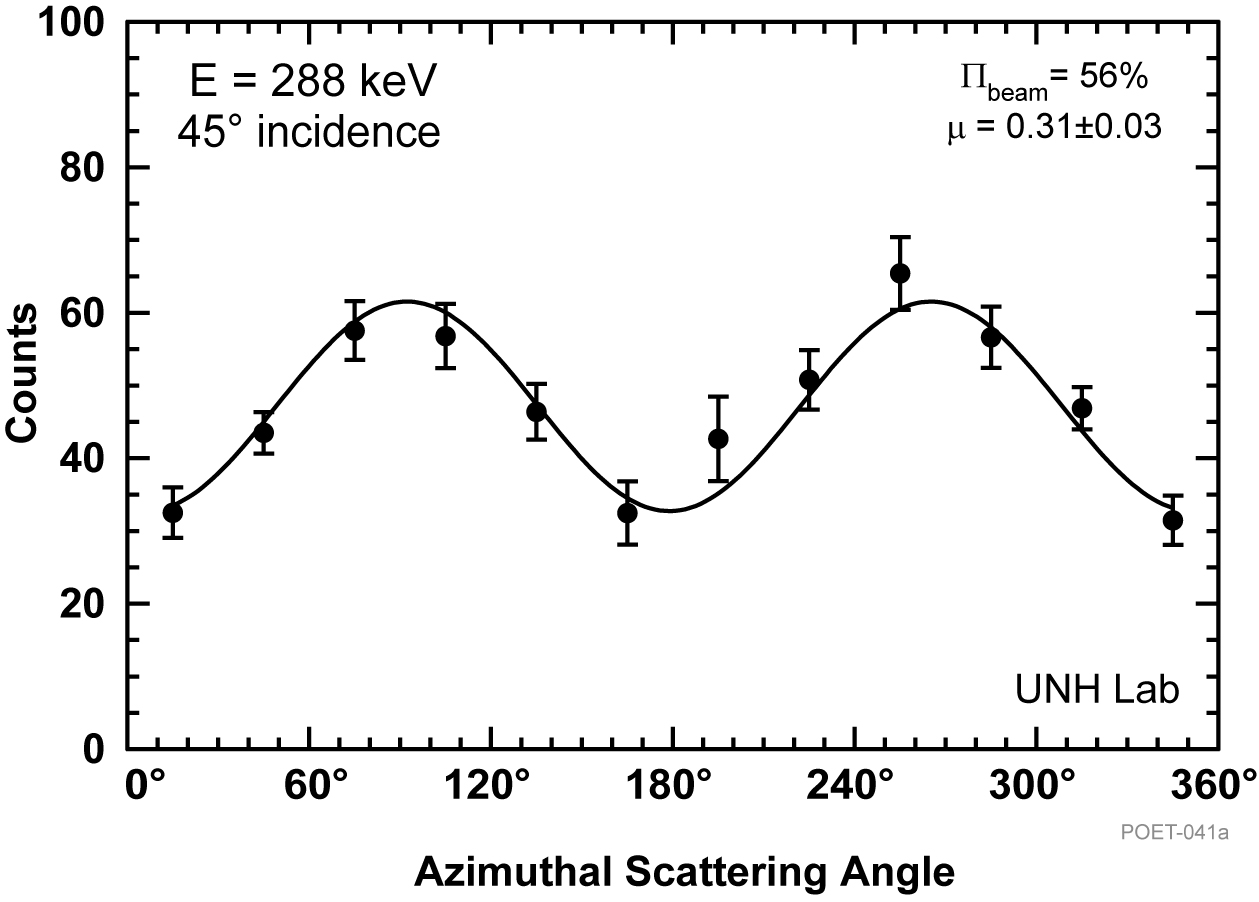}
     \includegraphics[height=0.2\textheight]{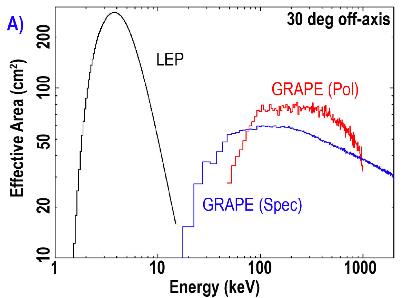}
  \caption{Left: GRAPE detection principle. Center: GRAPE off-axis response to 288 keV X-rays ($45^o$ off-axis). Right: POET Effective area $30^o$ off-axis}
\label{figGRAPE}
\end{figure}

\textbf{Polarimeter Modules:} The design of the GRAPE instrument is
very modular, with 62 independent polarimeter modules and 2
spectroscopy modules. Each polarimeter module incorporates
an array of optically independent 5x5x50 $mm^3$ non-hygroscopic
scintillator elements aligned with and optically coupled to the 8x8
scintillation light sensors of a 64-channel MAPMT. Two types of scintillators
are employed. Low-Z plastic scintillator is used as an effective
medium for Compton scattering. High-Z inorganic scintillator
(Bismuth Germanate, BGO) is used as a calorimeter, for absorbing
the full energy of the scattered photon. The arrangement of
scintillator elements within a module has 28 BGO calorimeter
elements surrounding 32 plastic scintillator scattering elements. Valid polarimeter
events are those in which a photon Compton scatters in one
of the plastic elements and is subsequently absorbed in one
of the BGO elements, as shown in Figure \ref{figGRAPE}. These events can
be identified as a coincident detection between one plastic
scintillator element and one BGO calorimeter element. The
azimuthal scatter angle is determined for each valid event by the
relative locations of hit scintillator elements. It is not necessary
to know where within each element the interaction takes place
(e.g., the depth of interaction). It is sufficient to know only the
lateral location of each element to generate a histogram of
photon scatter angles.
Each independent polarimeter module includes the electronics required to process the
MAPMT signals as well as qualify and digitize the event data.
This approach - compact, rugged, independent modules - was
employed successfully with two H8500 MAPMT-based detector
modules in the GRAPE balloon payload flown from Palestine,
TX in June 2007.  

\textbf{Spectrometer Modules:} To facilitate spectral measurements
over a broader energy range (15 keV--1 MeV), GRAPE
includes 2 spectrometer modules. Each
spectrometer module assembly has a single thallium-doped
sodium iodide (NaI(Tl)) scintillator crystal mounted on the face
of the MAPMT. Laboratory tests of this configuration
demonstrate energy resolution of $15\%$ and $8\%$ FWHM
at 122 and 662 keV, respectively.

\begin{description}
\item[Performance]
\end{description}
The scientific performance of the GRAPE polarimetry modules
has been investigated using a laboratory prototype module
and Monte Carlo simulations. The simulations were performed
using the software package MGEANT \citep{Stur03} with
a modification to include the effects of polarization in Compton
scattering \citep{McCon08}. The simulations were
validated with laboratory tests and have been used to predict
the performance of the GRAPE instrument on POET.

\textbf{Laboratory Testing:} The laboratory prototype is a configuration
of plastic scintillators and CsI crystals. The prototype
module was exposed to partially polarized radiation produced
by $90^o$ scattering of 662 keV photons (from a $^{137}Cs$ source) in
a plastic scintillator block, producing a beam with energy ~288
keV and a polarization of $\sim60\%$. The measured polarization
for normally-incident radiation was $56\%\pm 9\%$ was
consistent with the scattering geometry. Tests were also performed with an off-axis source to demonstrate the wide angle response of the GRAPE design at angles > $45^o$. The
response is shown in Figure \ref{figGRAPE}. 

\textbf{Beam Calibration:} The prototype module was exposed to $100\%$
polarized radiation of two energies at the Advance Photon Source at Argonne 
National Laboratory.  Modulation factors of $46\%$ (69 keV) and $48\%$ (129
keV), were obtained which are in agreement with simulations \citep{McCon08}.

\textbf{Engineering Balloon Flight:} In June 2007 the prototype
polarimeter module was flown on a NASA balloon flight from
Palestine, TX. The module performance met
all requirements throughout the flight. The measured
background counting rate (integrated over the full energy range
of the instrument) is $\sim2$ $cts$ $sec^{-1}$ per module.

\textbf{Simulations:} The full GRAPE instrument has been simulated
to predict its scientific performance. Both the effective area and modulation factor were
calculated using simulations of mono-energetic photons with
$100\%$ polarization. The off-axis effective area is shown as a function of energy in Figure \ref{figGRAPE}. Simulations show that GRAPE
retains $\sim40\%$ of its polarization sensitivity for photons incident
$60^o$ off-axis. Based on these performance simulations, we
expect GRAPE to detect $\sim40$ GRBs per year with a MDP better
than $20\%$, and $\sim6$ per year with MDP better than $8\%$.

\subsection{Low Energy Polarimeter: LEP}

The LEP measures the polarization of incident photons with
the innovative operation of a Time Projection Chamber (TPC), a proven technology used in
high-energy particle physics. It is simple in
design and uses the photoelectric effect to provide unmatched
broadband polarization sensitivity over the 2--15 keV band-pass
making it a low risk solution that requires low power and mass
to make highly sensitive measurements \citep{Black07}\citep{Hill07}.

\begin{figure}
  \includegraphics[height=.2\textheight]{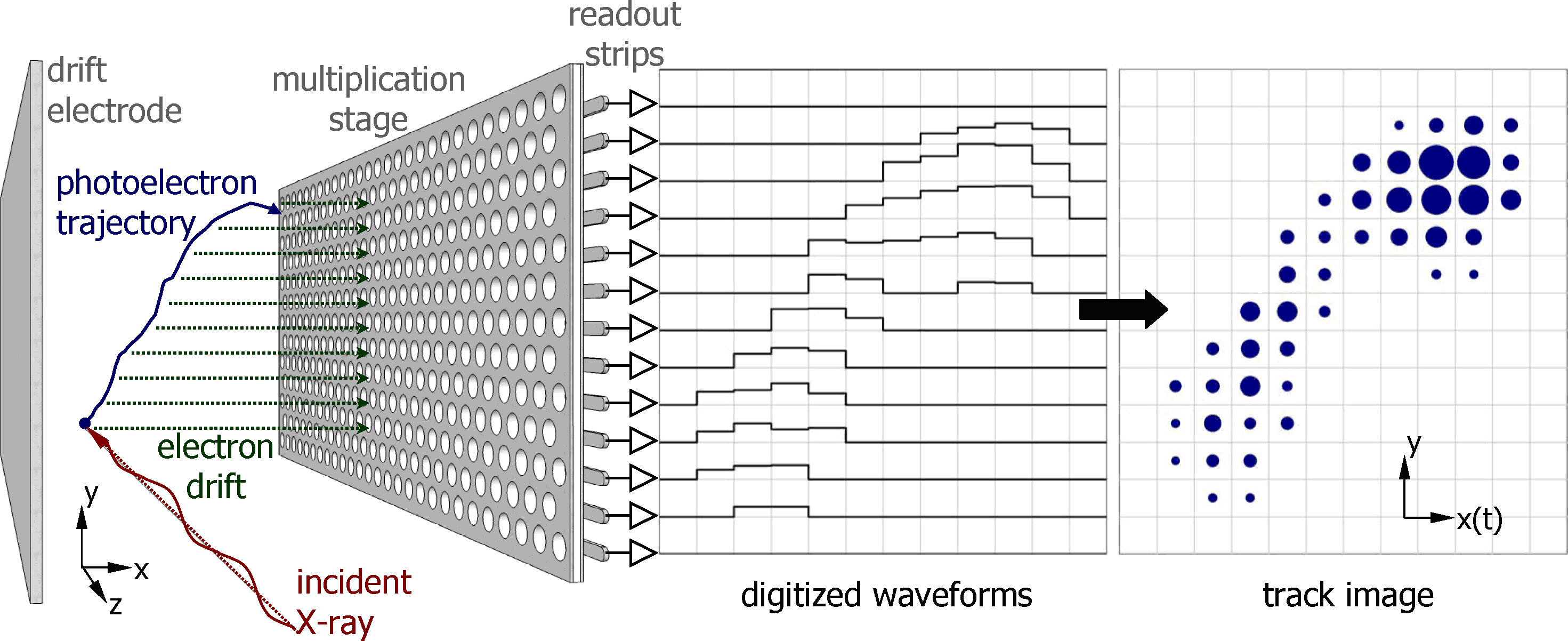}
    \includegraphics[height=.2\textheight]{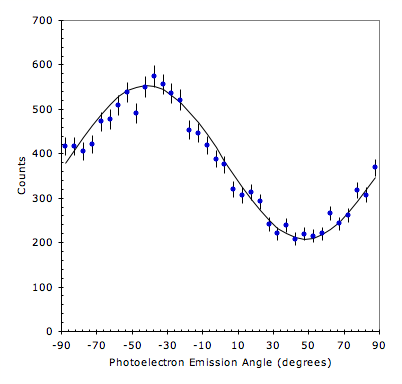}
  \caption{Left: The TPC polarimeter uses a simple strip readout and time of arrival to form a pixelized image of photoelectron track. Middle: The TPC polarimeter forms an image by digitizing the signal on each readout strip. The signal from a 6 keV X-ray, proportional to the charge pulse-train deposited on each strip, is shown on the right. The resulting image shows the interaction point, emission angle and end of the track. The size of each circle is proportional to the deposited charge in each virtual pixel; the grid is on a 132 $\mu$m spacing. Right: A histogram of reconstructed polarized 6.4 keV photoelectron tracks. Demonstrates a modulation of $45\%$.}
\label{figLEP}
\end{figure}

\textbf{TPC Polarimeter Operation:} The LEP polarimeter enclosure
will consist of four dual-readout detector modules each with an
isolated gas volume contained by a Be X-ray window. Each
detector module contains two 6 x 12 x 24 $cm^3$ (LxWxH) TPCs that
share a single X-ray transparent drift electrode. Each TPC is comprised of a micropattern
proportional counter, consisting of a shared drift electrode and
a high-field gas electron multiplier (GEM) positioned 1 mm from
a strip readout plane. When an X-ray is absorbed in the gas
between the drift electrode and the GEM, a photoelectron is
ejected in a preferential direction with a $cos^2\phi$ distribution, where
$\phi$ is the azimuthal angle measured from the X-ray polarization
vector. As the photoelectron travels through the gas it creates a
path of ionization that drifts in a moderate, uniform field to the
GEM where an avalanche occurs. The charge finally drifts to the
strip detector where it is read out.

Figure \ref{figLEP} illustrates how a track image projected onto
the x-y plane is formed by digitizing the charge pulse waveforms
and binning into pixels. The coordinates are defined by strip
location in one dimension, and arrival time multiplied by the drift
velocity in the orthogonal dimension. The strips are smaller than
the mean free path of the photoelectron and therefore an image
of the track can be reconstructed and the initial direction of the
photoelectron determined. The magnitude and orientation of the
source polarization can be determined from a histogram of the
emission angles.

\textbf{Design:} The LEP will use 12 x 24 $cm^2$ GEMs with holes in
a hexagonal configuration on an 80 $\mu$m pitch. The readout
plane is 12 x 24 $cm^2$, with strip electrodes 24 cm long on a 80
$\mu$m pitch and every $31^{st}$ strip tied together. The LEP will
use nitromethane ($CH_3NO_2$) as the charge-carrying ion in
neon, with a small quantity of $CO_2$ as a quench gas at 780 Torr
providing a slower drift speed compatible with slower, low power
electronics.

\textbf{Performance:}
A prototype detector has been
characterized with polarized X-rays at 6.4 keV and 4.5 keV and
unpolarized 6 keV X-rays in Ne:$CO_2$:$CH_3$$NO_2$ at 500 and 600
Torr. A typical modulation curve is shown in Figure \ref{figLEP}. The
modulation factor at 6.4 keV is $41\%$ and the residual instrumental
asymmetry measured with unpolarized X-rays is below $1\%$. 
The instrument performance has been modeled over the full energy
range. The LEP effective area for a source $30^o$ off-axis is shown
in Figure \ref{figGRAPE}, providing more than 200 $cm^2$ at 3.5 keV.

\textbf{GRB Sensitivity:} To determine the sensitivity of the LEP, GRBs
were selected from the HETE-2 catalog from the beginning of
the mission to 13 September 2003 which were detected both
in the wide-field X-ray monitor (WXM) and the French Gamma
telescope (FREGATE). 45 bursts satisfied these criteria
\citep{Sak05}.

The MDP for each of the 45 bursts was calculated using a
weighted modulation factor \citep{Pacc03}. Accounting
for the HETE-2 FoV (0.9 str) and the time period over which
the catalog was obtained ($\sim 3$ years operational time including
periods in the SAA), and scaling the distribution of bursts to the
LEP field of view, the number of bursts for a given MDP was
determined per year. 
LEP will detect >8 GRBs per year with a sensitivity of $10\%$ and >40/year with a sensitivity of $25\%$.

\subsection{POET Sensitivity}
\begin{figure}
  \includegraphics[height=.2\textheight]{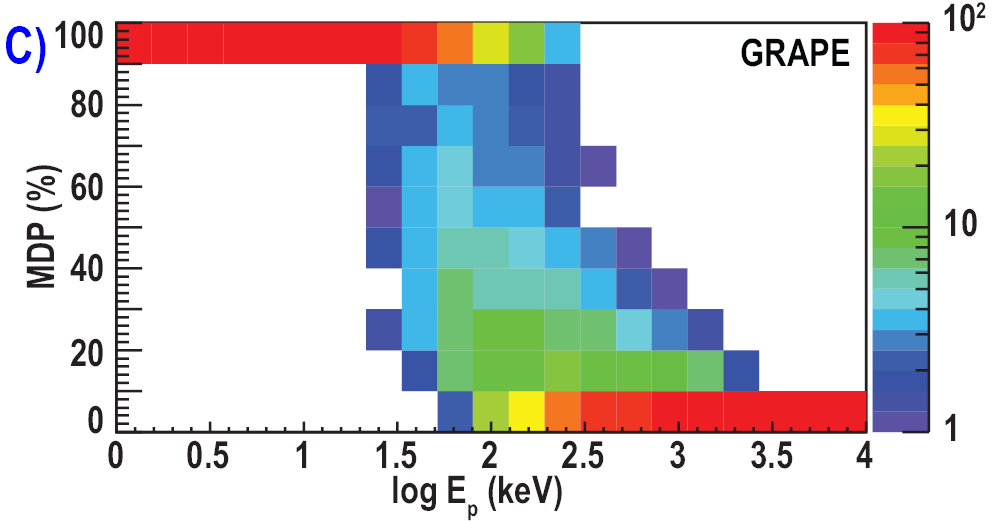}
   \includegraphics[height=.2\textheight]{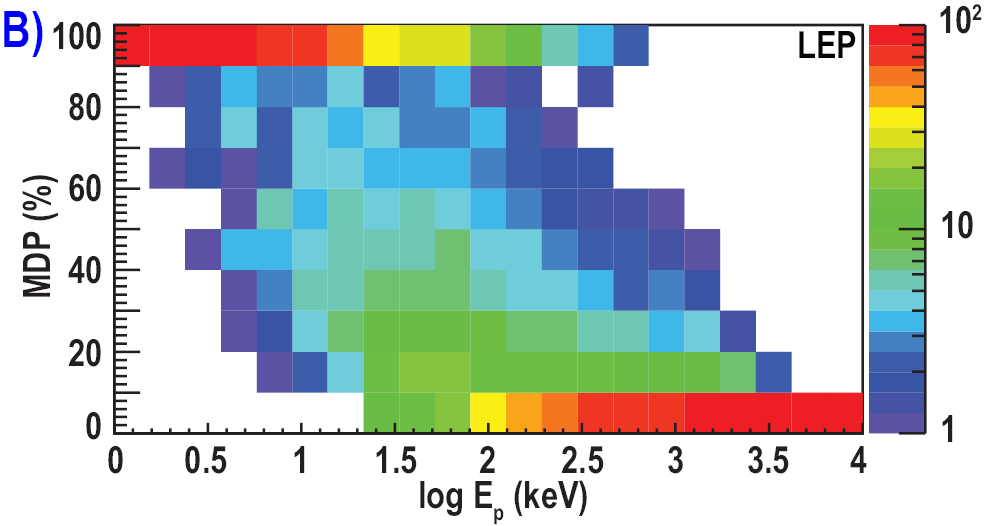}
  \caption{LEFT: GRAPE sensitivity to GRB polarization for a distribution of $E_p$  RIGHT: LEP sensitivity to GRB polarization for a distribution of $E_p$ }
  \label{figSens}
\end{figure}

44,100 GRBs were simulated, varying $E_p$ from
1 keV to 10 MeV. The BATSE time-averaged Band function
spectral parameter distribution was assumed for the low energy
and high-energy photon indices and the BAT $T_{100}$ distribution
for the duration. The time-averaged flux (2-400 keV) versus $E_p$
of the HETE-2 GRB sample \citep{Sak05} indicates that
GRBs with $E_p$ $\sim10$ keV, 100 keV and 1000 keV tend to have a
flux of $\sim10^{-8} erg$ $cm^{-2}$ $s^{-1}$, $\sim10^{-7} erg$ $cm^{-2}$ $s^{-1}$ and $\sim10^{-6} erg$ $cm^{-2}$ $s^{-1}$, respectively. The LEP and GRAPE responses were calculated
for the 12,500 simulated bursts that satisfied the HETE-2
relationship. The number of bursts detected with S/N>5 is $99\%$ for LEP, $80\%$ for GRAPE, $78\%$ GRAPE and
LEP). The number of bursts for which $E_p$ can be determined is
$20\%$ for $E_p$ < 10 keV, > $50\%$ for $E_p$ < 20 keV and $\sim100\%$ for $E_p$ $\sim0.2$
- 1 MeV. The polarization sensitivity (i.e. MDP) to the burst sample is shown in Figure \ref{figSens} where the z-axis is the percentage of bursts measured with a given $E_p$
and MDP.

\subsection{SUMMARY}
The capabilities presented here for GRAPE and LEP, show that the POET mission would significantly advance our understanding of key physical processes through high
energy polarimetry and simultaneous broadband spectroscopy of Gamma-ray bursts.


\begin{theacknowledgments}
The authors would like to acknowledge the efforts of the entire POET team that proposed to the 2008 SMEX AO and the USRA proposal team that compiled much of the information presented in this paper.
\end{theacknowledgments}



\bibliographystyle{aipprocl} 


\begin{thebibliography}{9}
\bibitem{Band93}
D. Band et al.,  \emph{ApJ} \textbf{413}, 281 (1993)

\bibitem{Black07}
J.~K. Black et al. \emph{NIM A}, \textbf{581}, 755 (2007)

\bibitem{FZP2005}
Y.~Z. Fan, B. Zhang \& D. Proga, \emph{ApJ} \textbf{635}, L129 (2005)

\bibitem{Gehrels05}
N. Gehrels et al., \emph{Nature} \textbf{437}, 851(2005)

\bibitem{Gran03}
J. Granot, \emph{ApJ} \textbf{596}, L17 (2003)

\bibitem{Hill07}
J.~E. Hill et al., \emph{Proc. SPIE} \textbf{6686}, 66860Y (2007)

\bibitem{Laz04}
D. Lazzati et al., \emph{MNRAS} \textbf{347}, L1 (2004)

\bibitem{LPB2003}
M. Lyutikov, V.~I. Pariev \& R.~D. Blandford, \emph{ApJ} \textbf{597}, 998 (2003)

\bibitem{McCon04}
M.~L. McConnell et al.,  \emph{Proc SPIE} \textbf{5165}, 334 (2004)

\bibitem{McCon08}
M.~L. McConnell et al., \emph{NIMA}, submitted 

\bibitem{Pacc03}
L. Pacciani et al., \emph{Proc SPIE}  \textbf{4843}, 394 (2003)

\bibitem{Pac1998}
B. Paczynski,  \emph{ApJ} \textbf{494}, L45 (1998)

\bibitem{RM1992}
M.~J. Rees \& P. Meszaros, \emph MNRAS \textbf{258}, 41P (1992)

\bibitem{RM1994}
M.~J. Rees \& P. Meszaros, \emph{ApJ} \textbf{430}, L93 (1994)

\bibitem{Ryde05}
F. Ryde, \emph{ApJ} \textbf{625}, L95 (2005)

\bibitem{Sak05}
T. Sakamoto et al., \emph{ApJ} \textbf{629}, 311 (2005)

\bibitem{SD1995}
N.~J. Shaviv and A. Dar, \emph{ApJ} \textbf{447}, 863 (1995)

\bibitem{Stur03}
S.~J. Sturner et al., \emph{A\&A} \textbf{411}, L81 (2003)

\bibitem{Toma08}
K. Toma et al., these proceedings

\bibitem{Wax03}
E. Waxman, \emph{Nature} \textbf{423}, 388 (2003)

\bibitem{ZM2004}
B. Zhang \& P. Meszaros, \emph{Int. J. Mod. Phys. A} \textbf{19}, 2385 (2004)


\end{thebibliography}

\IfFileExists{\jobname.bbl}{}
 {\typeout{}
  \typeout{******************************************}
  \typeout{** Please run "bibtex \jobname" to optain}
  \typeout{** the bibliography and then re-run LaTeX}
  \typeout{** twice to fix the references!}
  \typeout{******************************************}
  \typeout{}
 }


\end{document}